\documentstyle[12pt,epsfig]{article}

\def\ct#1{{\cal #1}}

\newcommand{\nn}{\nonumber}

\newcommand{\tp}{\tilde{\pi}}
\newcommand{\ts}{\tilde{\sigma}}
\newcommand{\tchi}{\tilde{\chi}}

\newcommand{\be}{\begin{equation}}
\newcommand{\ee}{\end{equation}}
\newcommand{\bea}{\begin{eqnarray}}
\newcommand{\eea}{\end{eqnarray}}

\begin{document}
\pagestyle{empty} 
\begin{flushright}
CERN-TH/97-266 \\
ROME1-1154/97 \\
ROM2F/97/35 \\
\today
\end{flushright}
\centerline{\LARGE \bf Towards $N=1$ Super-Yang--Mills}
\vskip 0.2cm
\centerline{\LARGE \bf on the Lattice}
\vskip 0.3cm
\centerline{\bf{A. Donini$^{a}$\footnote{Address after October 1$^{st}$: Universidad Autonoma de
Madrid, Cantoblanco, 28049 Madrid.}, M. Guagnelli$^b$\footnote{Address after October 1$^{st}$: 
Theory Group, DESY-IfH, Platanenallee 6, 15738 Zeuthen.}, 
P. Hernandez$^b$, A. Vladikas$^{c}$}}
\vskip 0.3cm
\centerline{$^a$ Dip. di Fisica, Univ. di Roma ``La Sapienza'' and
INFN, Sezione di Roma,}
\centerline{P.le A. Moro 2, I-00185 Roma, Italy.}
\smallskip
\centerline{$^b$ Theory Division, CERN, 1211 Geneva 23, Switzerland.}
\smallskip
\centerline{$^c$ INFN, Sezione di Roma II, and 
Dip. di Fisica, Univ. di Roma ``Tor Vergata'',}
\centerline{Via della Ricerca Scientifica 1, I-00133 Roma, Italy.}

\date{}
\abstract{
We consider the lattice regularization of $N=1$ supersymmetric
Yang--Mills theory with Wilson fermions. This formulation 
breaks supersymmetry at any finite lattice spacing; we discuss
how Ward identities
can be used to define a supersymmetric continuum limit, which 
coincides with the point where the gluino becomes massless. 
As a first step towards the understanding of the zero gluino--mass limit,
we present results on the quenched low-lying spectrum of SU(2) $N=1$
Super-Yang--Mills, at $\beta=2.6$ on a $V=16^3 \times 32$ lattice, in the
OZI approximation. Our results, in spite of the quenched and OZI
approximations, are in remarkable agreement with theoretical 
predictions in the supersymmetric theory, for the states with masses
which are not expected to get a large contribution from fermion loops.
}
\vskip 0.8cm
\begin{flushleft}
CERN-TH/97-266 \\
\today \\
\end{flushleft}
\vskip 0.5 cm
\vfill\eject
\pagestyle{empty}\clearpage
\setcounter{page}{1}
\pagestyle{plain}

\newpage
\pagestyle{plain} \setcounter{page}{1}

\section{Introduction}

Softly broken $N=1$ supersymmetry gives a natural solution to the 
long-standing hierarchy 
problem present in the Standard Model, considered as a low-energy 
effective model of a more fundamental theory that includes gravity.
In recent years, many new results on the non-perturbative regime of 
supersymmetric gauge theories have been derived and/or conjectured \cite{rev}. 
However, the only proved results have been obtained in the context of 
extended supersymmetry ($N > 1$), where the renormalization properties are 
severely constrained \cite{sw}. From these, rigorous results can also be 
derived in the phenomenologically more relevant cases of $N=1$ and $N=0$, 
through the introduction of explicit and {\it perturbative} soft-breaking 
terms \cite{others}.
Unfortunately, the analytical methods fail when the soft-breaking terms 
become large with respect to the scale of the strong dynamics. This is of course
the most interesting regime, and the lattice is a good non-perturbative 
method to explore it. 

In this work we will consider the case of $N=1$ Super-Yang--Mills theory (SYM).
Some preliminary results for gauge observables in this theory have been 
presented in \cite{m2} and confirmed with a different method in \cite{dg}.
The standard lore in SYM was conjectured a long time ago \cite{vy,rusians} and 
our programme is to confirm this picture on the lattice. It is 
believed that these theories show confinement and a mass gap. Furthermore SUSY 
is expected to be unbroken. This expectation is based on the fact that the Witten 
index is not zero \cite{witten}. As will be explained in this work, SUSY 
Ward identities (WIs)
can be used to study the issue of SUSY breaking on the lattice. At low energies, the relevant degrees of 
freedom are expected to be colourless composite fields of gluons
and gluinos and the lowest-lying spectrum 
can be described by a low-energy effective 
supersymmetric Lagrangian \cite{vy}. The bound 
states belong to chiral supermultiplets. The lowest-lying chiral
supermultiplet is expected to be \cite{vy}
\be
\Phi = ( \tp , \ts , \tchi , F , G ) ,
\label{eq:phi}
\ee
where $\tp = \bar \lambda \gamma_5 \lambda$ is a pseudoscalar,
$\ts = \bar \lambda \lambda$ is a scalar, and
$\tchi = \frac{1}{2} \sigma_{\mu\nu} F_{\mu\nu} \lambda$ is a fermionic
composite field (here $\lambda$ denotes the gluino field and $F_{\mu\nu}$
the gluon field tensor).
The fields $F$ and $G$ (scalar and pseudoscalar respectively) are
auxiliary fields with no dynamics. In the supersymmetric limit, these
particles should be degenerate in mass \cite{vy}. 
With supersymmetry softly broken by a gluino mass term, the mass
splittings between $\tp,\ts$ and $\tchi$ have been computed, to 
lowest order in the gluino mass, in  \cite{mv}. 

In this paper we will be concerned with supersymmetric $SU(2)$. 
Our lattice formulation, described in section 2, is simply the
Wilson action for the bosonic components and the Wilson--Dirac operator for 
the gluinos. The presence of Majorana fermions 
implies a modification of the standard measure of the path integral
and the rules of Wick contractions.
This lattice regularization breaks supersymmetry for two reasons. 
Firstly the breaking of Lorentz symmetry implies that the 
supersymmetric algebra cannot be satisfied at finite lattice spacing. Secondly, the Wilson 
term induces a finite mass for the gluino which splits the supermultiplet.
This problem is, however, not very different from the explicit breaking of chiral 
symmetry in lattice QCD with Wilson fermions  and can be handled in 
the same way \cite{cv}. The essential point is that the 
symmetry-breaking effects
that survive the continuum limit can be eliminated through renormalization
(if there are no SUSY 
anomalies), which in this case amounts to a proper tuning 
of the gluino bare mass. This follows from very general principles. Since the
lattice theory is asymptotically free, the continuum limit can be defined at 
the UV fixed point. In this limit the lattice theory must be equivalent to
a theory with the same degrees of freedom and all possible interactions of dimension four or less, 
which are compatible with the symmetries of the lattice action. In particular, from the exact gauge and cubic symmetries 
of the lattice action, it follows that the only possible
continuum action is the one corresponding to SYM, augmented by a gluino mass term. 
So even though SUSY is broken at finite lattice spacing in a very complicated
way, the continuum limit of this theory is softly broken SYM, provided there 
are no supersymmetric anomalies. Through a non-perturbative 
tuning of the bare gluino mass, a supersymmetric limit can be obtained, 
which coincides with the zero gluino-mass limit \cite{cv}. In section 3, 
we review how this can be achieved in practice through the use of Ward
identities. Moreover, we argue that an important simplification of this method
consists in the implementation of the OZI approximation. 
A method to determine the order parameter of supersymmetry breaking using
WI is also discussed in the same section. 

The low-lying spectrum defined in (\ref{eq:phi}) can be computed from
exponential decay of the two-point correlation functions at large time
separations, as described in section 4. The measurements in the scalar and
pseudoscalar channels are completely analogous to their QCD counterparts. The 
fermionic composite $\tchi$ is a new channel that presents difficulties, which
closely resemble those encountered in glueball spectroscopy in QCD and,
fortunately, can be overcome with the same type of techniques. 

In sections 5 and 6, we present our spectroscopy results
in the quenched and OZI approximations. Although SUSY is lost in these
approximations, this is a convenient first step, which enables lattice
calibrations and optimizations of the observables, without the high computer
cost of dynamical fermion simulations. It is also useful in the understanding
of the zero gluino-mass limit in a simpler setting, where the chiral anomaly
is switched off. As in quenched QCD, PCAC is expected to govern the chiral
restoration of the theory in the continuum limit. Our results strongly
support these expectations. Whenever comparison with the results of \cite{mv}
is possible, we find agreement within large statistical errors.
A preliminary version of our results can be found in \cite{dlatt97}.

\section{Lattice gauge theory with Majorana fermions}
\label{sec:act}

We use the lattice regularization of the $N=1$ SYM, first proposed in ref. \cite{cv}. The gauge field (the gluon), is the usual
link variable $U_\mu(x)$ in the fundamental representation of the gauge group 
$SU(N_c)$ (here $N_c=2$). The purely gluonic part of the action $S_g$
is the usual sum over plaquettes formed by link variables $U_\mu(x)$. 
The gluino $\lambda$ is in the adjoint 
representation of the gauge group and satisfies the Majorana condition
\bea
 \lambda = \lambda^{\ct{C}} = C \bar \lambda^T , \;\;\;\; 
\bar \lambda = \bar \lambda^{\ct{C}} = \lambda^T C ,
\label{eq:conjg}
\eea
where the superscript $\ct{C}$ stands for charge conjugation, the superscript
$T$ for transposition in spin and colour indices (implicit in the above
equations) and $C$ is the charge conjugation
matrix. 
The fermionic action is the Wilson--Dirac operator for adjoint fermions, i.e.
\be
S_f =  \sum_{x,y} \bar \lambda^a_\alpha(x)
\Delta^{ab}_{\alpha \beta}(x,y) \lambda^b_\beta(y) =  \sum_{x,y}
\lambda^a_\alpha(x) 
M^{ab}_{\alpha \beta}(x,y) \lambda^b_\beta(y) ,
\ee
where $M = C \Delta$ is the fermion matrix
\bea
M^{ab}_{\alpha \beta}(x,y) &=& \frac{1}{4} \sum_\mu V^{ab}_\mu(x) [C
(\gamma_\mu -r)]_{\alpha \beta} \delta^4_{y,x+\hat \mu} \nonumber \\
&-& \frac{1}{4} \sum_\mu V^{ab}_\mu(y) [C (\gamma_\mu +r)]_{\alpha \beta}
\delta^4_{y,x-\hat \mu}
\nonumber \\
&+& \frac{1}{2} (m_0 + 4r) \delta^{ab} C_{\alpha \beta} \delta^4_{y,x}  .
\eea
The basic notation and index conventions are given in Appendix \ref{app:conv}.
As usual, we have fixed $r=1$. 
The gauge field $V_\mu^{ab}(x)$ is defined by
\be
V_\mu^{ab}(x) = 2\; {\rm Tr}[U^\dagger_\mu(x) T^a U_\mu(x) T^b] ,
\ee
with the trace acting on colour indices in the fundamental representation.
It satisfies the properties \cite{m2}
\be
V_\mu^{ab}(x) = V_\mu^{ab}(x)^\ast = [V_\mu^{ab}(x)^T]^{-1} ,
\label{eq:realV}
\ee
where the transposition and inversion act on the adjoint colour indices $a,b$.
Using these properties, it can easily be shown that the field $V_\mu^{ab}(x)$
transforms in the adjoint representation of the colour group. 
The fermion matrix is antisymmetric:
\be
M^{ab}_{\alpha \beta}(x,y) = - M^{ba}_{\beta \alpha}(y,x) .
\ee
In the simulations, the fermion fields are rescaled, in standard
lattice QCD fashion, by a factor $\sqrt(1/2K)$, with the hopping parameter
given by $K = (2 m_0 + 8 r)^{-1}$.

The theory is defined by the path integral
\be
\ct{Z}[{J}] = \int \ct{D}U e^{-S_g} \int \ct{D}\lambda e^{- S_f -
\sum_x J(x) \lambda(x)} ,
\ee
where $J$ is an external source of the gluino field $\lambda$. Note
that the fermionic integration is only over $\ct{D}\lambda$ (unlike the
QCD case, where we integrate over $\ct{D}\lambda \ct{D}\bar \lambda$).
This is a consequence of the fact that, with Majorana fermions, the field
$\bar \lambda$ is not an independent degree of freedom; c.f.
eq.~(\ref{eq:conjg}). Since the action is at most quadratic in the fermion
fields, these can be integrated out by the standard change of variables:
\be
\lambda^a_\alpha (x)^\prime = \lambda^a_\alpha (x) + \frac{1}{2}
\sum_y J^b_\beta(y) [M^{-1}]_{\beta \alpha}^{ba} (y,x) \,\, ,
\ee
from which we obtain
\bea
\ct{Z}[J] &=& \int \ct{D}U e^{-S_g} \int \ct{D}\lambda^\prime 
e^{- \sum_{x,y} [ \lambda^\prime(x) M(x,y) \lambda^\prime(y) 
+ \frac{1}{4} J(x) M^{-1}(x,y) J(y) ] }
\nonumber \\
&=& \int \ct{D}U e^{-S_g} Pf[M] e^{- \frac{1}{4} \sum_{x,y}
J(x) M^{-1}(x,y) J(y)} ,
\eea
where $Pf[M]$ is the Pfaffian of the antisymmetric fermion matrix $M$
(see ref. \cite{m2}). The gluino propagator is defined as (using 
eq.~(\ref{eq:conjg}))
\bea
\label{eq:prop}
&& \langle T \{ \lambda(x) \bar \lambda(y) \} \rangle =
\langle T \{ \lambda(x) \lambda(y) \} \rangle C = \\
&& 2 \left [ \frac{\delta^2 \ln(Z[J])}{\delta J(x) \delta J(y)} \right ]_{J = 0} C =
\langle M^{-1}(x,y) \rangle C =
\langle \Delta^{-1}(x,y) \rangle , \nonumber
\eea
where $T\{...\}$ means time ordering. For the last term of the above equation,
we use the definition $M = C \Delta$ and the properties of the charge
conjugation matrix $C$; see Appendix \ref{app:conv}. An important consequence
of the Majorana nature of the gluino is that, unlike the case of Dirac fermions,
Wick contractions of the form $\lambda(x) \lambda(y) = M^{-1}(x,y)$ and
$\bar \lambda(x) \bar \lambda(y) = C^T M^{-1}(x,y) C$ are now allowed.

Note that our formulation is apparently different from that of \cite{m2},
where the action, path integral and gluino propagator are obtained in two
steps. First, the usual Wilson fermion action with Dirac fermions is expressed
in terms of Majorana fermions. Then, by using a ``doubling trick",  which
identifies ``half" the Dirac degrees of freedom with Majorana degrees of
freedom, all quantities of interest (such as the Majorana fermion action, path
integral and Green functions) are obtained. With this method (see
ref.~\cite{m2} for details) the gluino propagator is given by
\be
\langle T \{ \lambda(x) \bar \lambda(y) \}\rangle = \frac{1}{2} [
\langle \Delta^{-1}(x,y) \rangle + C^{-1} \langle(\Delta^{-1}(y,x))^T \rangle C]
\label{eq:double}
\ee
(here $T$ transposes spin and colour indices).
Using the antisymmetry of $M$ and the basic properties of $C$ (see Appendix
\ref{app:conv}), it can easily be shown that, on a single gauge field
configuration,
\be
\Delta(x,y) = C^{-1} \Delta (y,x)^T C ,
\label{eq:marco}
\ee
which implies that the two terms on the r.h.s. of eq.~(\ref{eq:double}) are
equal. Thus, the gluino propagator is identical in the two formalisms
(eqs.~(\ref{eq:prop}) and (\ref{eq:double})). Similar identities can easily be
derived for any correlation function of gluino fields. We therefore conclude
that the formalism of ref. \cite{m2} and the one presented here are equivalent; 
the latter is conceptually more straightforward, as it avoids the
``doubling trick".

We define the local composite operators that we will use in this work:
\bea
P(x) &=& \bar \lambda (x) \gamma_5 \lambda(x) ,\nn \\
S(x) &=& \bar \lambda (x) \lambda(x) \nn\\
A_{\mu}(x) & = & \bar \lambda(x) \gamma_\mu \gamma_5 \lambda(x),
\label{eq:oper} \\
\chi_\alpha(x) &=& \frac{1}{2} P^a_{\mu\nu}(x)
\left(\sigma_{\mu\nu}\right)_{\alpha\beta} \lambda^a_\beta, \nn
\eea
where $P_{\mu\nu}$ is the lattice transcription of the field strength $F_{\mu\nu}$,
\be
\label{eq:pmunu}
P_{\mu\nu} \equiv \frac{U_{\mu\nu} - U^\dagger_{\mu\nu} }{2 i g_0},
\label{pmunu}
\ee
with $g_0$ the bare lattice coupling and $U_{\mu\nu}$ the Clover
open plaquette.
The reason for this choice is that $P_{\mu\nu}$ transforms under parity and
time-reversal in the same way as $F_{\mu\nu}$ in the continuum; see
Appendix \ref{app:discr}. Finally, its dual tensor is defined as
\begin{equation}
  \label{dual}
  \tilde P_{\mu\nu}(x) = \frac{1}{2} \epsilon_{\mu\nu\rho\sigma} 
P_{\rho\sigma}(x).
\end{equation}

\section{Chiral and supersymmetric Ward identities}
\label{sec:wis}

The action described in the previous section is not supersymmetric. However,
as explained in the introduction, this can be cured in the continuum limit
through an appropriate tuning of the bare gluino mass (and of operator mixing 
coefficients, if composite operators are involved) \cite{cv}.
This is completely analogous to the problem of restoration of non-singlet 
chiral symmetries in lattice QCD, with Wilson fermions, best formulated
through WIs \cite{boc}. In subsec.~\ref{sub:susyres}
we review the mechanism of SUSY restoration on the lattice and show how the
gluino mass can be tuned to the supersymmetric limit non-perturbatively
with the aid of axial and/or SUSY WIs. This procedure may be difficult to
implement in practice. For
this reason, we argue in subsec.~\ref{sub:ozi} that the tuning of the gluino mass
can also be realized in the much simpler framework of axial WIs in the OZI
approximation. Finally, in subsec.~\ref{sub:op} we discuss the determination
of the order parameter of SUSY breaking from SUSY WIs. Unlike the rest of this
paper, all quantities in this section are in physical (rather than lattice)
units.

\subsection{Ward identities and restoration of supersymmetry}
\label{sub:susyres}

For any global symmetry which is broken by the lattice regularization, the
bare WIs have a form which differs from their continuum counterparts by
additional terms arising from the explicit breaking. 
However, the renormalization conditions can be fixed by the requirement that 
the renormalized WIs have the same form as the continuum ones, up to terms
which vanish in the continuum limit. In our case, there are two global
symmetries at the classical level: the first is the symmetry under a chiral
rotation of the gluino field, which is broken by the anomaly. The second is
supersymmetry. On the lattice both are broken explicitly. Under axial rotations
the following WI is obtained
\be
\nabla_{\mu} A_{\mu} = 2 m_0 P + X_A, 
\label{cwib}
\ee
where $P$ and $A_{\mu}$ are defined in eq.~(\ref{eq:oper}), $m_0$ is the bare
gluino mass and $X_A$ is a dimension-five operator with two fermion legs 
\footnote{The explicit expression for $X_A$ can be found in \cite{cv}.}. 

All quantities in the above
relation are bare and need renormalization. Both $A_{\mu}$ and $P$ are 
renormalized multiplicatively, whereas $X_A$ can mix with several operators
of lower dimension. By using the symmetries of the action, namely parity,
gauge invariance and cubic invariance, three such operators are identified,
in terms of which a finite ${\bar X}_A$ can be defined as
\be
\bar{X}_A = X_A + ( Z_A - 1 ) \nabla_\mu A_{\mu} - {\tilde Z}_A \nabla_{\mu} A_{\mu} 
- Z_Q P_{\mu\nu}{\tilde P_{\mu\nu}} + 2 {\bar m} P .
\label{eq:xa}
\ee
$P_{\mu\nu}$ is defined in eq.~(\ref{pmunu}) and $\tilde P_{\mu\nu}$ in
eq.~(\ref{dual}).
The renormalization condition is chosen to be the vanishing of the matrix
element of $\bar{X}_A$ between gluino or physical (colourless) states in the
continuum limit. Schematically we write $\lim_{a\rightarrow 0}\;
\bar{X}_A  = 0$. This implies that the renormalized WI is 
\be
 \nabla_{\mu} {\hat A}_{\mu} = 2 (m_0-{\bar m}) Z^{-1}_P {\hat P} + {\hat Q} 
+ \ct{O}(a), 
\label{cwir}
\ee
where
\bea
\hat A_{\mu} &=& Z_A A_{\mu}, \nonumber \\
\hat P &=& Z_P P , \\
\hat Q &=& Z_Q P_{\mu\nu}{\tilde P_{\mu\nu}}+ {\tilde Z}_A \nabla_{\mu}
A_{\mu} . \nonumber
\eea
Note that the renormalization of the anomalous term $\hat Q$ is not
multiplicative. This fact was anticipated in eq.~(\ref{eq:xa}), where the
mixing with $\nabla_\mu A_\mu$ was artificially split into two terms, 
in order to highlight that one term (proportional to $Z_A-1$) renormalizes
the current while the other (proportional to $\tilde Z_A$) renormalizes
the operator $P_{\mu\nu}\tilde P_{\mu\nu}$. This WI has the same form as the
continuum one, provided we identify the renormalized gluino mass with 
\be
\hat m_{\lambda} =  (m_0 - \bar{m}) Z^{-1}_P.
\label{mass1}
\ee
For details see the second paper of ref.\cite{unga}.

For the supersymmetric WI, the procedure is analogous \cite{cv}. Performing
supersymmetric transformations on the gluino and gauge fields (see
Appendix \ref{app:xaxs}), we obtain the bare lattice WI
\be
\nabla_{\mu} S_{\mu} = 2 m_0 \chi + X_S ,
\label{swib}
\ee
where $S_{\mu}(x)$ is the lattice version of the supersymmetric current,
$\chi$ is defined in eq.~(\ref{eq:oper}) and $X_S$ is a sum of operators of
dimension 11/2. The explicit expressions of $S_\mu$ and $X_S$ has been worked
out in Appendix \ref{app:xaxs}.

As before, we have to renormalize this relation. The identification of all
operators of equal or lower dimension,
with the same quantum numbers \cite{cv} defines a finite (subtracted)
$\bar{X}_S$. Imposing the renormalization condition (schematically)
$\lim_{a\rightarrow 0}\; \bar{X}_S  = 0$, we obtain the renormalized WI
\be
\nabla_{\mu} {\hat S}_{\mu} = 2 (m_0 -\tilde{m}) Z^{-1}_{\chi} \hat \chi + 
\ct{O}(a),   
\label{swir}
\ee
where
\bea
\hat \chi &=& Z_\chi \chi, \\
\hat S_{\mu} = Z_S S_{\mu} + Z_T T_{\mu}, \nonumber
\eea
with $T_{\mu}(x)\equiv \gamma_{\nu} P^a_{\nu\mu}(x) \lambda^a(x)$. This has
the same form as the continuum WI, provided we identify the renormalized
gluino mass with
\be
\hat m_{\lambda} =  (m_0 - \tilde{m}) Z^{-1}_{\chi}.
\label{mass2}
\ee

The compatibility of (\ref{mass1}) and (\ref{mass2}) implies
\be
(m_0-\tilde{m}) Z^{-1}_{\chi} = (m_0-\bar{m}) Z^{-1}_{P}.
\ee
Both sides of the equation are functions of $m_0$ and, for fixed $g_0$, and they
 vanish at the same critical value $m_0=m_{cr}$, as expected. The supersymmetric 
limit is restored when
\be
m_0 = m_{cr} = \tilde m(m_{cr},g_0) = \bar m(m_{cr},g_0). 
\ee

In principle, the WIs~(\ref{cwir}) and (\ref{swir}) can be used for
an explicit non-perturbative verification of the above scenario.
The supersymmetric WI (\ref{swir}) for two independent operators $O_{i}$,
becomes
\bea
\frac{Z_S}{Z_T} \langle \nabla_{\mu} S_{\mu}(x)  O_i(0)  \rangle + \langle \nabla_{\mu} T_{\mu}(x)  
O_i(0) \rangle = 2 \frac{(m_0 - {\tilde m})}{Z_T} 
\langle \chi(x) O_i(0), \rangle 
\label{diff}
\eea
with $i=1,2$. This is valid for $x \ne 0$ or for $O_i$ invariant under SUSY
transformations (otherwise there would also be contact terms).
The above two equations (for $i=1,2$) can be solved for
$Z_S/Z_T$ and $(m_0 - {\tilde m})/Z_T$.
Similarly, from the axial WI~(\ref{cwir}), we obtain
\bea
\frac{Z_A-\tilde Z_A}{Z_Q} \langle \nabla_{\mu} A_{\mu}(x)  O_i(0)  \rangle -
\langle P_{\mu\nu}(x) \tilde P_{\mu\nu}(x)
O_i(0) \rangle = 2 \frac{(m_0 - {\bar m})}{Z_Q} 
\langle P(x) O_i(0) \rangle, 
\label{diffch}
\eea
which can be solved for $(Z_A-\tilde Z_A)/Z_Q$ and $(m_0 -
{\bar m})/Z_Q$. Thus, at finite UV cutoff $a^{-1}$ the ratios
of renormalization constants can be determined and, more importantly,
the tuning of the gluino mass to its critical value
can be performed in two independent
ways. However, since at least two independent amplitudes for each WI
insertion are needed to tune $m_0$, this might be difficult in practice.

\subsection{Ward identities in the OZI approximation}
\label{sub:ozi}

\begin{figure}[t]   
\vspace{0.1cm}
\centerline{\epsfig{figure=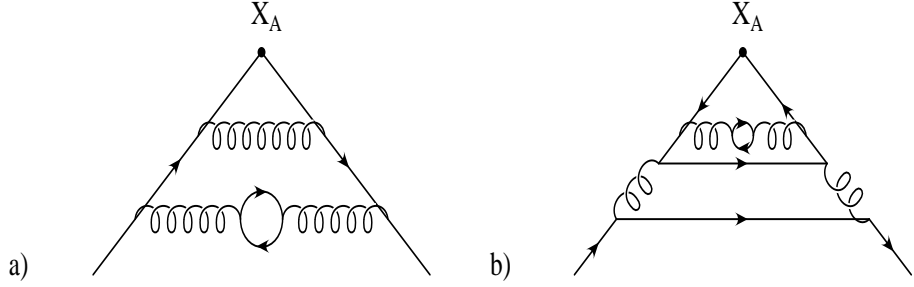,height=4cm,width=12cm,angle=0}}
\caption{Examples of the two types of insertion of $X_A$: (a) OZI diagram; (b) Non-OZI diagram.} 
\label{fig:diag}
\end{figure}
In order to obtain a simple alternative method for the tuning of the
gluino mass to its critical value, we consider the axial WI in the ``OZI
approximation''. It consists in separating
the insertions of the operator $X_A$, appearing in any correlation function,
 into 
two contributions: those arising from diagrams in which the two fermion
fields of $X_A$ are not contracted (OZI) and those arising from
diagrams in which they are (non-OZI); see Fig. \ref{fig:diag}. 
Note that
the anomaly operator $P_{\mu\nu}\tilde P_{\mu\nu}$ can only arise as a
counterterm to the non-OZI contributions. Thus, we may render the OZI
part of $X_A$ finite by defining the subtraction
\be
\bar X_{A,OZI} = \left[ X_A + (Z_A^\prime - 1) \nabla_\mu A_\mu +
2 \bar m^\prime P \right]_{OZI}
\label{eq:subozi}
\ee
and imposing $\lim_{a\rightarrow 0}\; \bar{X}_{A,OZI}  = 0$.
A renormalized WI can then be obtained
\be
 \nabla_{\mu} {\hat A}^\prime_{\mu} =
2 (m_0-{\bar m}^\prime) (Z_P^\prime)^{-1} {\hat P}^\prime  + 
\ct{O}(a), 
\label{cwiozi}
\ee
with ${\hat A}^\prime_{\mu} = Z^\prime_A A_{\mu}$ and $\hat P^\prime
= Z^\prime_P P$.  
This WI is an operator relation, which holds in any OZI amplitude.
Now, using eq.~(\ref{cwiozi}) it is easy to get the point 
$m_0 = \bar{m}'(m_0,g_0)$ by tuning to zero the ratio $2 \rho_{OZI}$:
\be
2 \rho_{OZI} \equiv \frac{\nabla_{\mu} \langle {A}_{\mu}(x) P(0)
\rangle|_{OZI}}{\langle P(x) 
P(0) \rangle|_{OZI}} = \frac{m_0 - {\bar m}'}{Z'_A}.
\label{eq:rhoozi}
\ee
However, this is only useful if $m_0- {\bar m}^\prime$ vanishes at the same 
critical point where $m_0 - {\bar m} = 0$, which is not at all 
obvious. To see how this may come about, we also consider the
subtractions generated by the non-OZI part of $X_A$; these are
\be
\bar X_{A,non-OZI} = \left[ X_A - Z_A^{\prime\prime} \nabla_\mu A_\mu 
- Z_Q P_{\mu\nu} \tilde P_{\mu\nu} + 2 \bar m^{\prime \prime} P
\right]_{non-OZI}.
\ee
The sum of OZI and non-OZI insertions must reproduce the anomalous
renormalized WI (\ref{cwir}). This implies that the OZI tuning of
$m_0- {\bar m}^\prime$ and that of $m_0- {\bar m}$ occur at the same point
$m_{cr}$, provided 
\be
{\bar m}^{\prime \prime}(m_{cr},g_0^2) = 0 .
\label{eq:ass}
\ee
This is expected to be the case for a theory with more 
gluino flavours, $N_f > 1$. In such a theory, the previous reasoning holds, but
in addition there is an
additional non-singlet chiral symmetry, which yields the standard
non-anomalous PCAC WI. The explicit breaking of this symmetry, contained in $X^{NS}_A$, 
can be subtracted through renormalization
\be
\bar X^{NS}_{A} = X^{NS}_A + (Z_A^\prime - 1) \nabla_\mu A^{NS}_\mu +
2 \bar m^\prime P^{NS} ,
\ee  
where the superscript $NS$ stands for non-singlet current and operator.
As usual, the renormalization condition is $\lim_{a\rightarrow 0} \bar X^{NS}_{A} = 0$. 
Notice that $Z'_A$ and $\bar m'$ are identical to those appearing in the 
OZI singlet WI, eq. (\ref{eq:subozi}), since the insertions of $X^{NS}_A$ 
only have OZI contributions.  

In the continuum limit we should recover the expected continuum WIs:
\bea
 \nabla_{\mu} {\hat A}_{\mu}^{NS} &=& 2 \hat m_\lambda {\hat P}^{NS}, \\
 \nabla_{\mu} {\hat A}_{\mu} &=& 2 \hat m_\lambda {\hat P} + {\hat Q}. \nonumber
\label{cwirns}
\eea
Notice that, in the continuum, the renormalized gluino mass appearing in both WIs is the same.
It then follows that
\be
\hat m_\lambda = \frac{m_0 - \bar m^\prime - \bar m^{\prime \prime} }{Z_P} = \frac{m_0 
- \bar m^\prime}{Z_P^{NS}},
\ee
where $\hat P^{NS} = Z_P^{NS} P^{NS}$. This implies that, at the critical point defined from 
eq. (\ref{eq:rhoozi}), eq. (\ref{eq:ass}) is verified. 
If we further assume that the dependence of $\bar m''(m_{cr}, g_0)$  on $N_f$ is
analytic,  this result should also hold  in the $N_f = 1$ case. In this case, 
the point of 
symmetry restoration is given, up to $\ct{O}(a)$ terms, by the vanishing of
$2\rho_{OZI}$; c.f. eq.(\ref{eq:rhoozi}).

The computations of the present paper have been performed in the quenched-OZI
approximation. In this case, the requirement that $\bar m''$ be analytic in $N_f$
can be relaxed, since the OZI two-point correlation 
functions are identical to the non-singlet ones in a theory with $N_f = 2$ 
(internal fermion loops, which could depend on $N_f$, are absent). 
Hence, the tuning to the zero gluino-mass limit can be done using eq. (\ref{eq:rhoozi}).  
Furthermore, it is believed that, in these approximations,
spontaneous chiral symmetry breaking occurs because of a non-vanishing gluino
condensate \cite{vy,rusians}.
If that is the case, the OZI pseudoscalar $\tp$ should become massless in the zero
gluino--mass limit. The two requirements $m_{\tp} \rightarrow 0$ and
$2 \rho \rightarrow 0$ 
give two determinations of $m_0$, which should only coincide if chiral
symmetry is 
spontaneously broken. We will see that this is confirmed numerically.
  
\subsection{Order parameter of supersymmetry breaking}
\label{sub:op}

WIs can also be used to study the issue of spontaneous 
supersymmetry breaking, in complete analogy with the study of 
spontaneous breaking of the non-singlet chiral symmetry in QCD described in \cite{boc}. 
Consider the supersymmetric WI,
\be
\langle \nabla_{\mu} {\hat S}_{\mu}(x) \hat \chi(0) \rangle =
2 (m_0 - \tilde m ) Z^{-1}_\chi \langle {\hat \chi}(x) 
\hat \chi(0) \rangle + \;\delta^4(x) \langle {\hat F}(x) \rangle .
\ee
The operator $\hat F$ is the renormalized 
vacuum expectation value of the bare auxiliary field
of the $\Phi$ supermultiplet. Thus it is the order 
parameter of SUSY breaking. For a non-zero gluino mass, we can 
integrate the previous relation over space-time and obtain
\be
0 = 2 \;(m_0 - \tilde m ) \sum_x Z^{-1}_\chi \langle \hat \chi(x) \hat 
\chi(0) \rangle + \langle \hat F(0) \rangle.
\ee
The renormalized order parameter for spontaneous SUSY breaking can then 
be computed as
\be
\langle \hat F(0) \rangle = - 2\; \lim_{m_0 \rightarrow m_{cr}} \;\; (m_0 - \tilde m) \sum_x Z^{-1}_\chi 
\langle \hat \chi(x) \hat \chi(0) \rangle.
\ee
If this expectation value does not vanish, then SUSY is spontaneously broken.
In this case the WI also implies that, in the supersymmetric limit, 
there exists a 
massless fermion field which propagates in the $\langle \chi \chi \rangle$ channel. 
This is the Goldstone fermion of supersymmetry breaking 
(the goldstino). Hence, another signal of spontaneous supersymmetry 
breaking would be to find a massless fermion in the channel $\langle\chi(x) \chi(y)\rangle$
in the limit of zero gluino mass (where this limit is defined  by using 
eq.~(\ref{diff}) or eq.~(\ref{diffch})). 

\section{Correlation functions}
\label{sec:corr}

In this section we discuss the method for extracting the lowest-lying
masses of the pseudoscalar, scalar and spinor particles of the supermultiplet
$\Phi$ (i.e. $\tp$, $\ts$ and $\tchi$, respectively), in the OZI approximation.

The two-point correlation functions used to obtain the 
$\tp$ and $\ts$  masses are
\bea
C_{PP}(t) &=& \sum_{\vec x} \langle P(\vec x,t) P(\vec x_0,t_0) \rangle ,
\nn \\
C_{SS}(t)  &=& \sum_{\vec x} \langle S(\vec x,t) S(\vec x_0,t_0) \rangle .
\label{eq:corrps}
\eea
In order to compute $2\rho$ we need also the following correlation functions:
\bea
C_{AP}(t)  &=&  \sum_{\vec x} \langle A_0(\vec x,t) P(\vec x_0,t_0) \rangle ,
\nn \\
C_{PA}(t)  &=&  \sum_{\vec x} \langle P(\vec x,t) A_0(\vec x_0,t_0) \rangle .
\label{eq:corrap}
\eea
In the computation of all these correlations, the point ($\vec x_0,t_0$) 
is held fixed at the origin. The correlations $C_{AP}$ and $C_{PA}$ are
antisymmetrized, in order to enhance the signal-to-noise ratio (from
now on $C_{AP}$ stands for the antisymmetrized correlation function).
The quantity $2\rho$ is obtained from the ratio:
\be
R(t) = \frac{\nabla_0 C_{AP}(t)}{C_{PP}(t)} .
\label{eq:rat}
\ee
where $\nabla_\mu f(x) \equiv [ f(x+\hat \mu) - f(x-\hat \mu) ] /2$ is a
symmetric lattice derivative.

We now express the two-point correlation functions of
eqs.~(\ref{eq:corrps})--(\ref{eq:rat}) in terms of gluino propagators.
The correlation function 
\be
C(t) = \sum_{\vec x} \langle O_1(x) O_2(x_0) \rangle
\ee
of two generic bilocal operators
$O_i = \bar \lambda \Gamma_i \lambda$, with $\Gamma_i$ any
Dirac matrix ($i=1,2$), can be written as
\bea
C(t) =
&-& \sum_{\vec x} \langle {\rm Tr} [Q(\vec x,t; \vec x_0, t_0; U_\mu) \Gamma_2
Q(\vec x_0, t_0; \vec x,t ; U_\mu) \Gamma_1 ] \rangle
\nonumber \\
&+& \sum_{\vec x} \langle {\rm Tr} [Q(\vec x,t; \vec x, t; U_\mu) \Gamma_1 ]
{\rm Tr} [ Q(\vec x_0, t_0; \vec x_0,t_0 ; U_\mu) \Gamma_2 ] \rangle \\
&+& \sum_{\vec x} \langle {\rm Tr} [Q(\vec x,t; \vec x_0, t_0; U_\mu) \Gamma_2
Q(\vec x_0, t_0; \vec x,t ; U_\mu) C \Gamma_1^T C ] \rangle ,
\nonumber
\eea
where $Q(\vec x, t ; \vec 0, 0; U_\mu) =
\Delta^{-1} (\vec x, t ; \vec 0, 0; U_\mu)$ is the fermion propagator,
computed on a single background field configuration (i.e. the inverse of the
fermion matrix). The first two terms of the above expression are the familiar
one- and two-boundary quark diagrams which also appear in the Dirac fermion
formulation. The third term arises from the extra contractions
characteristic of Majorana fermions; see also comments after
eq.~(\ref{eq:prop}). It is also a one-boundary quark diagram. For the
cases of interest (i.e. eqs.~(\ref{eq:corrps})--(\ref{eq:rat})), we have that
$C \Gamma_1^T C = -\Gamma_1$, and the first and third terms are equal.
In the present work we are interested in the OZI
approximation of the above expression, for which the second term does not
contribute. In other words, we compute only one-boundary diagrams. 

For the $\tchi$ mass, we use the two-point correlation function
\bea
[C_{\tchi}(t)]_{\alpha \beta} &=& \sum_{\vec x,\vec x_0}
\langle \chi_\alpha(\vec x,t) \bar \chi_\beta(\vec x_0,t_0) \rangle \nn \\
&=&\frac{1}{4} \langle
\left(\sigma_{\mu\nu}\right)_{\alpha\alpha'} P_{\mu\nu}^a(\vec x,t)
Q^{ab}_{\alpha'\beta'}(\vec x, t ; \vec x_0, t_0 ; U_\mu)
P_{\rho\sigma}^b(\vec x_0,t_0)\left(\sigma_{\rho\sigma}\right)_{\beta'\beta}
\rangle .
\eea
This correlation remains unchanged in the OZI approximation, as it only has
one valence gluino line. For $C_{\tchi}(t)$, we keep $t_0$ at the origin and
vary $\vec x_0$ over the whole time-slice $t_0$. This is obtained in the
simulations by solving numerically the equation
\be
\Delta^{aa'}_{\alpha\beta'}(\vec x, t ; \vec x', t')
\Sigma^{a'}_{\beta'\beta}(\vec x', t' ; \vec x_0, t_0)
= \delta^3_{\vec x , \vec x_0} \delta_{t,t_0}
P_{\mu\nu}^a(\vec x,t)\left(\sigma_{\mu\nu}\right)_{\alpha\beta} .
\label{eq:invchi}
\ee
The solution of the above is
\be
\Sigma^{a}_{\alpha\beta}(\vec x, t ; \vec x_0, t_0) =
Q^{aa'}_{\alpha\alpha'}(\vec x, t; \vec x_0, t_0; U_\mu)
P_{\mu\nu}^{a'}(\vec x_0,t_0)\left(\sigma_{\mu\nu}\right)_{\alpha'\beta},
\label{eq:solchi}
\ee
as can be verified by back substitution. Once $\Sigma$ has been obtained
numerically by inverting eq.~(\ref{eq:invchi}), the correlation $C_{\tchi}$
can be easily computed. By using various properties (parity, time reversal, etc.)
we show in Appendix \ref{app:discr} that the correlation function has the form
\be
[C_{\tchi}(t)]_{\alpha \beta} = C_1(t) \delta_{\alpha\beta} +
C_2(t) \gamma^0_{\alpha\beta} .
\label{eq:chiff}
\ee
We recall that the local operator $\chi_\alpha(x)$ also extends 
in the time direction, since it contains the field strength tensor 
$P^a_{\mu\nu}$. In order to eliminate the contact terms arising from 
adjacent time-slices, we have also considered the operator that has 
only the spatial component of the field strength, defined as follows:
\be
\chi^S_\alpha(x) = \frac{1}{2} F^a_{ij}(x)
\left(\sigma_{ij}\right)_{\alpha\beta} \lambda^a_\beta,
\ee
where $i,j=1,2,3$.
Its two-point correlation function is given by:
\bea
[C^S_{\tchi}(t)]_{\alpha \beta} &=& \sum_{\vec x,\vec x_0}
\langle \chi^S_\alpha(\vec x,t) \bar \chi^S_\beta(\vec x_0,t_0) \rangle \nn \\
&=&\frac{1}{4}
\left(\sigma_{ij}\right)_{\alpha\alpha'} P_{ij}^a(\vec x,t)
Q^{ab}_{\alpha'\beta'}(\vec x, t ; \vec x_0, t_0)
P_{kl}^b(\vec x_0,t_0)\left(\sigma_{kl}\right)_{\beta'\beta}, 
\label{eq:chispa}
\eea
where $i,j,k,l = 1,2,3$.

The $\tp$, $\ts$ and $\tchi$ masses are extracted from the asymptotic
behaviour of the corresponding correlation functions at large time separations.
In principle, both form factors of eq.~(\ref{eq:chiff}) should decay
exponentially in time with the same mass. Hence, we can obtain in principle
two independent measurements of the $\tchi$ mass. 

\section{Particle masses}
\label{sec:mass}
 
We have performed simulations at $\beta = 2.6$.
The lattice volume is $V = 16^3 \times 32$. 
We have generated quenched configurations, separated by 
400 thermalizing sweeps, with a standard hybrid heat-bath algorithm.
We have measured 950 gluino propagators (the measurement has been made every 30 sweeps) at
four values of the hopping parameter: $K = 0.174,\, 0.178,\, 0.182,\, 0.184$. 
In \cite{km}, simulations performed at $\beta =2.3$ gave too large a value
for the $\ts$ (greater than 1.5 in lattice units).
We have opted for a larger $\beta$ value, hoping to
keep all masses smaller than 1 (in lattice units), by
reducing the UV cutoff $a$.
Statistical errors have been obtained with the jacknife method, by
discarding 10 configurations at a time.
Unfortunately the time extension of the lattice seems a little
too short to isolate the lowest-lying state in the $\tp$ channel 
and two-exponential fits are needed in this case.
This increases the statistical error in this channel.

\subsection{The $\tp$ particle}
\label{subsec:tp}

We isolate the lowest-lying mass of the correlation function $C_{PP}(t)$
by fitting it with the function
\be
C_{PP}(t) = \frac{Z_1}{2m} \exp(-mt) \left [ 1 + Z_2 \exp(-\Delta mt) \right]
+ \left ( t \rightarrow T-t \right ) .
\label{eq:fit2m}
\ee
We have used this two-state fit in order
to estimate the lowest mass, by fitting in time interval $t_i \leq t \leq T/2 - 1$,
where $T = 32$ is the time extension of the lattice.
In table \ref{tab:fitmassp1} we show the results of our two-state fits
at the different values of the gluino mass as a function of the initial time
$t_i$.
\begin{table}
\centering
\begin{tabular}{|r|r|r|r|r|r|r|}
\hline
$K$ & $t_i=6$ & $t_i=7$ & $t_i=8$ & $t_i=9$ & $t_i=10$ \\
\hline \hline
$0.174$ & $0.869(2)$ & $0.867(2)$ & $0.864(5)$ & $0.855(6)$ & $0.860(7)$ \\
$0.174$ & $ - $ & $ - $ & $ 0.876(1) $ & $ - $ & $ - $ \\ \hline
$0.178$ & $0.713(2)$ & $0.711(2)$ & $0.707(5)$ & $0.698(8)$ & $0.703(11)$ \\
$0.178$ & $ - $ & $ - $ & $ 0.720(1) $ & $ - $ & $ - $ \\ \hline
$0.182$ & $0.536(2)$ & $0.534(3)$ & $0.531(4)$ & $0.526(11)$ & $0.528(11)$ \\
$0.182$ & $ - $ & $ - $ & $ 0.543(1) $ & $ - $ & $ - $ \\ \hline
$0.184$ & $0.433(2)$ & $0.432(3)$ & $0.430(4)$ & $0.428(7) $ & $0.426(9)$ \\
$0.184$ & $ - $ & $ - $ & $ 0.441(1) $ & $ - $ & $ - $ \\
\hline
\end{tabular}
\caption{The mass of $\tp$ as a function of initial
fitting time. For each $K$ value, data at the top (for $6 \leq t_i \leq 10)$
are the results from two-mass fits, whereas the data at the bottom of the
$t_i = 8$ column are those from one-mass fits.}
\label{tab:fitmassp1}
\end{table}

We note that the two-mass fit results decrease with increasing $t_i$
in the interval $6 \leq t_i \leq 8$.
This probably signals the weak residual presence of a higher mass state.
Since, however, from $t_i = 8$ onwards, the values are fluctuating within
their error, the isolation of two low-lying states is achieved.
As $t_i$ increases, so does the statistical error. We take as the best
estimates of the $\tp$ mass  the results of $t_i=8$.
As a further test, we did a one-state fit, also shown in table
\ref{tab:fitmassp1}, in the interval  $8 \leq t \leq 16$. The results
are systematically higher than the ones obtained at the same interval
with the two-mass fit. This further justifies the necessity for the
latter fit.

\subsection{The $\ts$ particle}
\label{subsec:ts}

The $\ts$ particle is noisier than the $\tp$. 
In table \ref{tab:fitmasss1} we show the results of the following
one-state fit for this correlation:
\be
C_{SS}(t) = Z_0 + \frac{Z_1}{2m} \left[ \exp\left( m(t-T)\right) +
\exp(-mt) \right] .
\label{eq:fits}
\ee
We take as our best estimates
the masses from $t_i = 10$. In all cases, the constant $Z_0$ was compatible
with zero (and orders of magnitude smaller than $Z_1$). This is hardly
surprising, since, working in the OZI approximation, we are only measuring the
contribution of the one-boundary diagram to the above correlation function.
This contribution is equal to the two-point correlation function of the
corresponding non-singlet scalar operator $S^{NS}(x)$, for which 
$Z_0 = 0$.

\begin{figure}[t]   
\vspace{0.1cm}
\centerline{\epsfig{figure=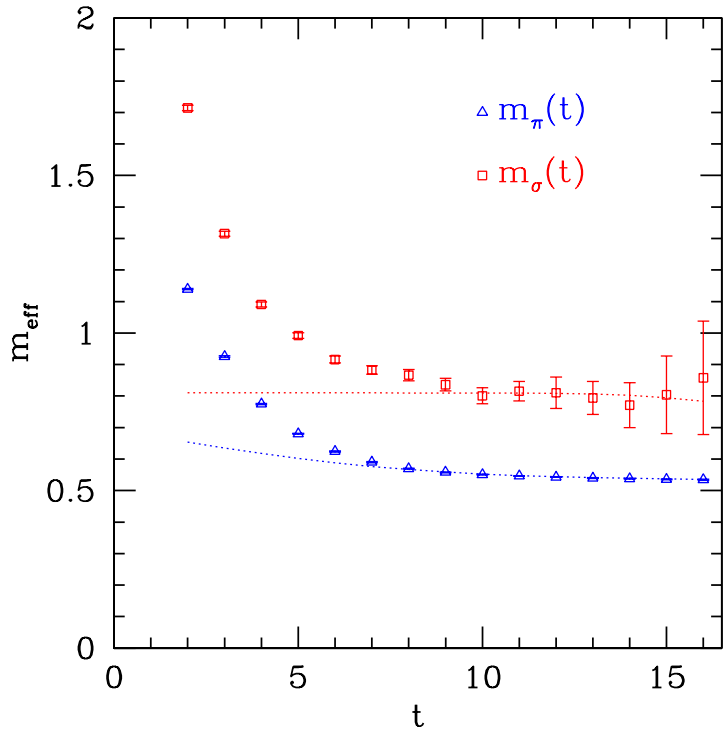,height=10cm,angle=0}}
\caption{Effective masses for $\tp$ (below) and $\ts$ (above) as a 
function of time at $K=0.182$, compared with results of the 
fits (dotted lines).} 
\label{fig:em_ps}
\end{figure}

In fig.~\ref{fig:em_ps} we present the effective masses for the
$\tp$ and $\ts$ as a function of $t$, compared with the results of 
eqs.~(\ref{eq:fit2m}) and (\ref{eq:fits}) at $K=0.182$. 
It is clear that the curves fit the data well. However,
lattices with larger time extensions are needed for the effective mass
of $\tp$ to reach a plateau.
In the case of the $\ts$, the effective mass plot displays a plateau.
The straightforward measurement of the $\ts$ mass in our case must be
contrasted to the situation in quenched QCD, where the signal 
for the scalar particle $\sigma$ is extremely noisy.

\begin{table}
\centering
\begin{tabular}{|r|r|r|r|r|r|r|}
\hline
$K$ & $t_i=7$ & $t_i=8$ & $t_i=9$ & $t_i=10$ & $t_i=11$ \\
\hline \hline
$0.174$ & $1.13(1)$ & $1.11(2)$ & $1.10(2)$ & $1.09(2)$ & $1.08(3)$ \\
$0.178$ & $0.99(1)$ & $0.97(2)$ & $0.95(2)$ & $0.95(3)$ & $0.93(3)$ \\
$0.182$ & $0.83(2)$ & $0.82(2)$ & $0.81(3)$ & $0.81(4)$ & $0.80(5)$ \\
$0.184$ & $0.75(3)$ & $0.74(3)$ & $0.74(5)$ & $0.76(6)$ & $0.76(9)$ \\
\hline
\end{tabular}
\caption{The mass of $\ts$ as a function of initial fitting time.}
\label{tab:fitmasss1} 
\end{table}

\subsection{The $2\rho$ correlation}
\label{subsec:2rho}

\begin{figure}[t]   
\vspace{0.1cm}
\centerline{\epsfig{figure=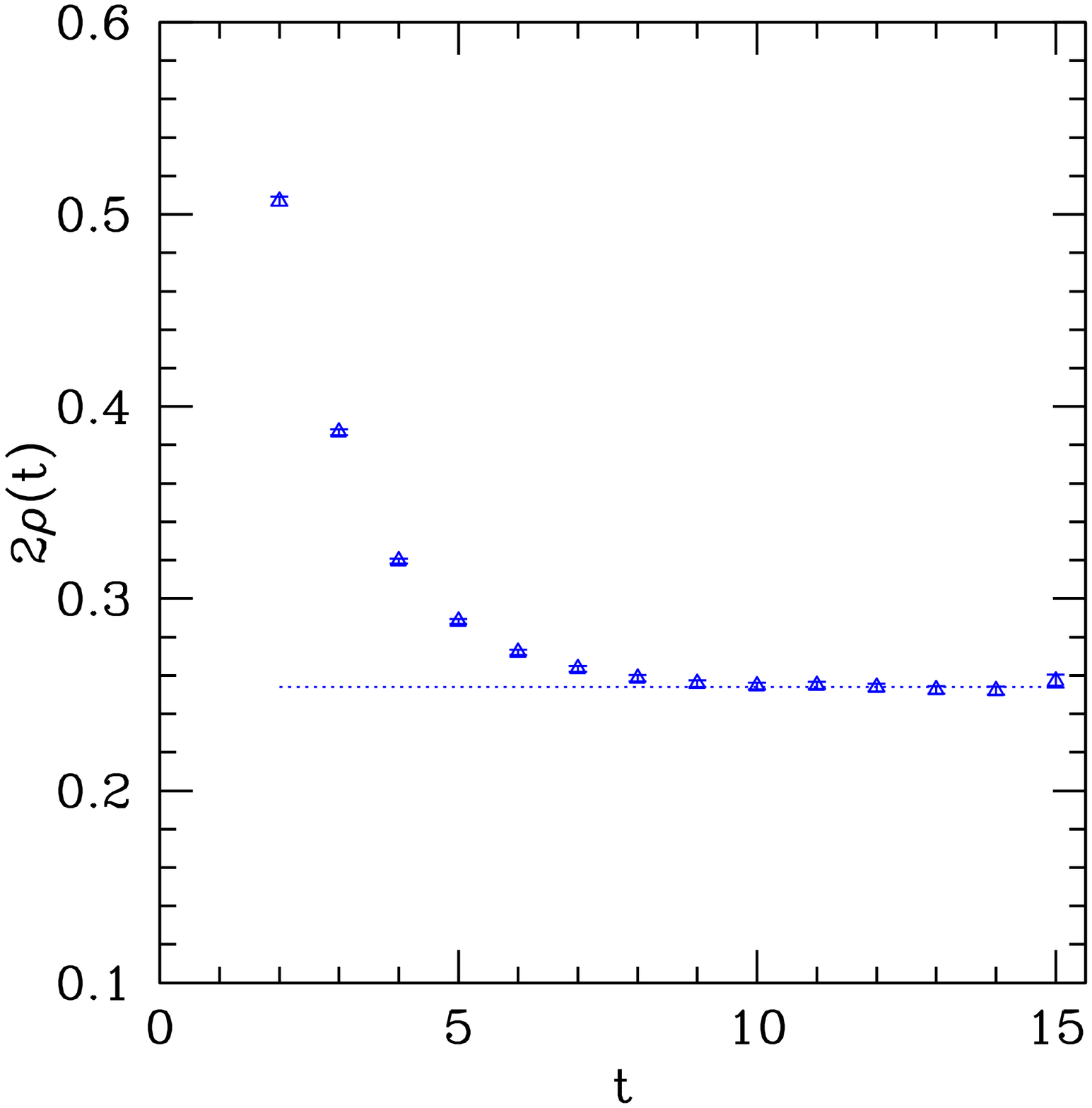,height=10cm,angle=0}}
\caption{$R(t)$ as a function of time at $K=0.178$. The dotted line is its average value
in the time interval 9-14.}
\label{fig:2rho}
\end{figure}

The ratio $R(t)$, which gives the estimate of $2\rho$, is shown in
fig.~\ref{fig:2rho} (for $K=0.178$). A plateau sets in at fairly early times. 
We perform a weighted average of the results in the time
interval $9 \leq t \leq 14$. The $2\rho$ estimates thus obtained are shown
in table \ref{tab:fit2rho}.
\begin{table}
\centering
\begin{tabular}{|r|r|r|r|r|}
\hline
$K$     & $0.174$   & $0.178$   & $0.182$   & $0.184$ \\ \hline\hline
$2\rho$ & $0.379(2)$ & $0.254(2)$ & $0.145(1)$ & $0.096(1)$\\
\hline
\end{tabular}
\caption{The values of $2\rho$ as a function of $K$.}
\label{tab:fit2rho}
\end{table}

\subsection{The $\tchi$ particle}
\label{subsec:chi}

\begin{figure}[t]   
\vspace{0.1cm}
\centerline{\epsfig{figure=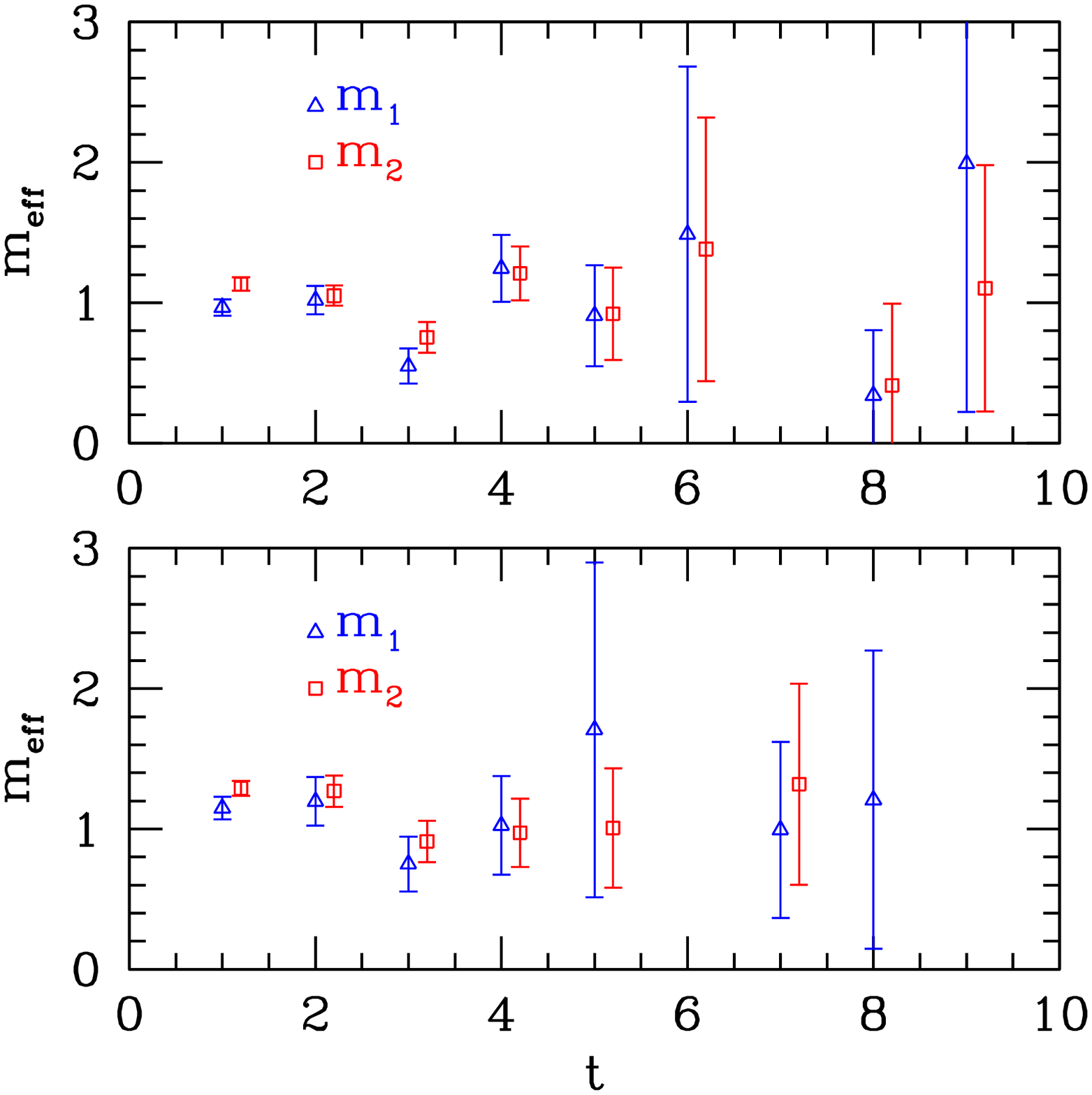,height=10cm,angle=0}}
\caption{Effective mass of $C_1(t)$ and $C_2(t)$ at $K=0.174$
as a function of time ($m_1$ and $m_2$ respectively in the text), 
with $n_s=4$ (below) and $n_s=8$ (above).}
\label{fig:chi_sme}
\end{figure}

We have obtained the effective mass of $\tchi$ from the spatial correlation
function of eq.~(\ref{eq:chispa}). We used all 
the properties of $C_{\tchi}^S(t)$ under the discrete symmetries
(explained in full detail in Appendix B), in order to isolate the two
independent components of the correlation, namely $C_1(t)$ 
and $C_2(t)$. No effective mass can be obtained naively from either $C_1(t)$
or $C_2(t)$, owing to large statistical fluctuations. In order to
improve the signal-to-noise ratio, we implemented the smearing {\it \`a la}
APE  \cite{APE_sme}, which is widely adopted in glueball spectroscopy.
The value of the smearing optimization parameter $\epsilon$ of \cite{APE_sme}
is fixed at $\epsilon = 0.5$. We have obtained results with two
smearing steps, namely $n_s = 4$ and $n_s=8$. In fig.~\ref{fig:chi_sme}
we show the effective mass extracted from $C_1(t)$ and $C_2(t)$
at $K=0.174$ as a function of time ($m_1$ and $m_2$ respectively), with $n_s=4$ and $n_s=8$. 
The results obtained with $n_s = 4$ show a plateau at small
times; this is reminiscent of the results for the glueball masses,
obtained with the same method. The results with $n_s = 8$ have somewhat
smaller statistical errors up to times $t=5$, but the values fluctuate
more as a function of $t$. We plan to study the optimization of the number
of smearing steps in a future simulation. Since the signal obtained from
$C_2(t)$ is cleaner than that obtained from  $C_1(t)$, in this
paper we use the former correlation function in order to extract the $\tchi$ mass.
We take as our best estimates of the $\tchi$ mass the results for $m_2$
in the interval $3 \le t \le 5$, for $n_s=4$, shown in table \ref{tab:fit_chi}.
\begin{table}
\centering
\begin{tabular}{|r|r|r|r|r|}
\hline
$K$     & $0.174$   & $0.178$   & $0.182$   & $0.184$ \\ \hline\hline
$m_{\tchi}$ & $0.93(12)$ & $0.84(11)$ & $0.73(10)$ & $0.66(9)$\\
\hline
\end{tabular}
\caption{The values of the $\tchi$ effective mass as a function of $K$.}
\label{tab:fit_chi}
\end{table}

\section{The chiral limit}
\label{sec:chiral}

As explained in section 3, in the quenched OZI approximation, chiral symmetry 
is not broken by the anomaly, but it is expected to be broken spontaneously \cite{vy}.
If this is the case, there are two numerically independent determinations of 
the critical $m_0(K_c)$ at which the gluino mass vanishes. The first one is obtained 
upon fitting the $\tp$ mass as a function of the hopping parameter
according to the standard PCAC relation:
\be
m_{\tp}^2 = A \left( \frac{1}{K} - \frac{1}{K_c} \right).
\ee
By using this relation we find the chiral limit at
\be
K_c = 0.18752(9)
\ee
(with one-mass fits, we obtain $K_c = 0.18759(2)$). 
The second possibility is to extract $K_c$ from the relation 
between two-point correlation functions derived from the WI (\ref{eq:rhoozi}) 
in this approximation, which implies the vanishing
of $2 \rho$. Extrapolating linearly $2\rho$ in $1/K$ we obtain
\be
K_c = 0.18762(4),
\ee
which is compatible with the previous estimate. The behaviour of the
OZI component of $\tp$ is thus similar to that of the $\pi$ particle in QCD, 
and consistent with the expectation of spontaneous chiral symmetry breaking.  
A discrepancy in the two estimates of $K_c$
would  imply, either that chiral symmetry is not broken (in which case the 
$\tp$ does not need to be massless for zero gluino mass, while the WI still 
has to be satisfied) or, more
probably, the presence of large $\ct{O}(a)$ discretization
errors in the correlation functions\footnote{For a recent exhaustive study of 
discretization effects in QCD, see ref. \cite{clv}.}. We see that none of these
possibilities occur.

\begin{figure}[t]   
\vspace{0.1cm}
\centerline{\epsfig{figure=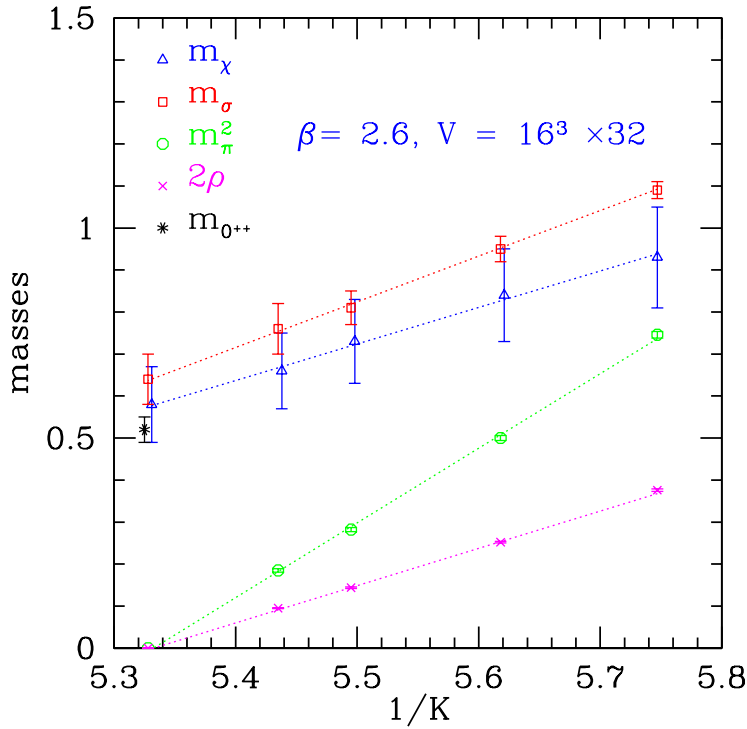,height=10cm,angle=0}}
\caption{Masses of $\tp,\,\ts,\,\tchi$ and $2\rho$ as a function of
$1/K$ (the mass of $\tp$ is squared). The extrapolation at $K_c$ is also shown (curves).
The value of the $0^{++}$ glueball mass at $\beta = 2.6$ is also included.}
\label{fig:mass}
\end{figure}

In order to obtain the masses of $\tchi$ and $\ts$ at $K_c$, we 
have performed a linear extrapolation in $1/K$  (from the
results of the fit for $\ts$ and from the effective mass for $\tchi$).
The results are shown in fig.~\ref{fig:mass}. 
It is clear that $\ts$ and $\tchi$ do not vanish at $K_c$. This 
demonstrates that the model has QCD-like
chiral behaviour, in accordance with the expectations of refs. \cite{vy,cv}. 
The non-vanishing of $\ts$ at $K_c$ had already been
seen in \cite{km}, at $\beta = 2.3$. A comparison with \cite{km} is 
possible through the ratio of the $\ts$ mass to the lowest glueball mass ($0^{++}$), which
has been measured for both $\beta = 2.3$ and $\beta = 2.6$ in \cite{mt}. The result is 
$\frac{m_{\ts}}{m_{0^{++}}} = 1.23(16)$ at $\beta = 2.6$ 
and $1.35(16)$ at $\beta = 2.3$. The two results are in fair agreement. 

The dependence of the scalar and fermion masses on the gluino mass is
compatible with linearity. This is expected to be the case in the 
unquenched theory, where the splittings in the multiplet
have been computed to lowest order in the gluino mass \cite{mv}:
\bea
m_{\tp}& = &C  + \frac{5}{ 6 } C' m_\lambda , \nn\\
\label{masve}
m_{\ts}& = &C  + \frac{7}{ 6 } C' m_\lambda ,  \\
m_{\tchi}& = &C  +  C' m_\lambda . \nn
\eea
$C$ and $C'$ are constants that in general depend on the renormalization 
scheme
\footnote{Note that there is disagreement on the results of the splitting
between the two papers of ref.\cite{mv}. In rederiving eq.~(\ref{masve}),
we found agreement with the second paper of \cite{mv}.}.
We can compare the ratio of the slopes with the theoretical 
prediction $\Delta_{th} =  1.17$. We find
\be
\Delta = \frac{d m_{\ts}}{d m_{\tchi}} = 1.3(5) .
\ee
Although 
the error in the slope ratio is quite large, the agreement is encouraging. 
Moreover, the masses of $\ts$ and $\tchi$ appear to be degenerate at $K_c$. 
Of course, this does not have to be the case in the quenched-OZI approximation. 
Nevertheless, the mass of the $\tp$ particle is expected to change 
dramatically because of the anomaly, whereas
there are no reasons to expect a dramatic change of
the $\ts$ and $\tchi$ masses when the OZI and quenched approximations are relaxed. 
In fig. \ref{fig:mass} the $0^{++}$ glueball mass at $\beta = 2.6$ is also included.
We recall that this particle represents an auxiliary field in the $\Phi$ supermultiplet,
and it should be degenerate in mass with the other particles belonging to $\Phi$
in the supersymmetric limit. Its mass appears to be compatible with the extrapolated 
$\tchi$ mass and $2\sigma$ below the $\ts$ mass. This small discrepancy 
could of course be due to the effect of the
OZI and/or quenched approximations on the $\ts$
mass (in particular,
notice that the OZI approximation only affects the $\ts$ correlation function, 
but neither $\tchi$ nor the glueball). 

\section{Conclusions}

We have measured the low-lying spectrum of $SU(2)$ with one flavour of adjoint matter. 
This is the matter content of supersymmetric $SU(2)$. Our
quenched simulation, although not yet supersymmetric, is a first step 
towards the understanding of the fine-tuning to the zero gluino--mass limit, 
which is necessary in the lattice formulation of $N=1$ supersymmetry 
with Wilson fermions. We have used the chiral WI to restore this limit 
in this work and discussed how this can also be done in the full theory. 

The numerical results on the spectrum are quite encouraging. In the full 
theory the theoretical expectation is that SUSY is not broken spontaneously 
and the low-lying states are the components of a composite chiral superfield, 
i.e. one complex spin zero particle and its fermionic partner. 
Of these fields, only the imaginary part of the scalar field gets its mass from 
fermion loops, through the chiral anomaly. 
The other states are not expected to change very much 
when the quenched approximation is relaxed. In fact, we have found that 
these states are approximately degenerate at zero gluino mass, 
already in the quenched approximation. This could be a first signal of
supersymmetry restoration at zero gluino mass on the lattice. 
Finally, the vanishing of the $\tp$ mass at the critical point   
is an indication of spontaneous chiral symmetry breaking, while the finite
$\tchi$ mass supports the theoretical expectation that SUSY
is not spontaneously broken (although this result must be confirmed in
the framework of unquenched simulations).

\begin{center}
{\bf Acknowledgements}
\end{center}
We thank M.B.~Gavela, K.~Jansen, I.~Montvay,
G.C.~Rossi, R.~Sommer, M.~Testa and G.~Veneziano 
for encouragement and/or many helpful discussions. 
We also thank G.~Martinelli and the Rome-I lattice group
for providing the computing resources that have made this work possible.

\appendix 
\section{Notation and conventions}
\label{app:conv}

Here we list our conventions.
The number of colours is $N_c$ (in this work $N_c=2$).
The number of flavours is $N_f$ (in this work $N_f=1$).
Latin lower-case indices (of the form
$a,b,\ldots = 1,\ldots, N_c^2-1$) stand for colour
in the adjoint representation.
Latin upper-case indices stand for colour in the fundamental
representation ($A,B,\ldots = 1,\ldots,N_c$).
Greek lower case indices from the beginning of the
alphabet stand for spin ($\alpha,\beta,\ldots = 1,\ldots,4$)
and those from the middle of the alphabet ($\mu,\nu,\ldots = 0,\ldots,3$)
denote Lorentz space-time components (in Euclidean space-time). Latin
lower-case indices (of the form $i,j,k,l = 1,2,3$) denote Lorentz
spatial components. 

The gauge field is thus $U^{AB}_\mu(x)$,
the fermion field is $\lambda_\alpha^a(x)$, the field-strength tensor
is $F_{\mu\nu}^a(x)$, etc. It is also customary to write $\lambda_\alpha(x) =
\lambda_\alpha^a(x) T^a$, $F_{\mu\nu}(x) = F_{\mu\nu}^a(x)T^a$,
with the colour indices in the fundamental representation (e.g.
$\lambda_\alpha^{AB}$, $T^a_{AB}$ etc.) supressed.
$T^a$ are the group generators, normalized as follows:
\be
{\rm Tr}[T^a T^b] = \frac{1}{2} \delta^{ab} .
\ee
For $N_c=2$, these are the standard Pauli matrices $T^a = \sigma^a/2$.

The Dirac matrices are in the Weyl representation in Euclidean space-time.
In terms of the three Pauli spin matrices and the unit $2 \times 2$ matrix
$I$, they are given by
\begin{equation}
\begin{array}{c} 
\gamma_0
\end{array}
=
\left(\begin{array}{rr}
 0 & I \\
 I & 0
\end{array}\right);
\qquad
\begin{array}{c} 
\gamma_k
\end{array}
= -i
\left(\begin{array}{rr}
 0 & \sigma_k \\
 -\sigma_k & 0
\end{array}\right) .
\label{eq:gammas}
\end{equation}
We also define
\bea
\gamma_5 &=& \gamma_1 \gamma_2 \gamma_3 \gamma_0 , \nonumber \\
\sigma_{\mu\nu} &=& \frac{1}{2} [ \gamma_\mu, \gamma_\nu] , \\
C &=& \gamma_0 \gamma_2.\nonumber
\eea
We often make use of the following property of the charge conjugation matrix:
\be
C^{-1} = -C = C^T .
\ee
Finally, we take the sign convention $\epsilon_{0123} = 1$.

\section{Discrete symmetries on the lattice}
\label{app:discr}

In the spirit of  ref. \cite{b}, we list the discrete symmetry transformations
of the gauge field $U_\mu(x)$ and the fermion fields $\lambda(x)$ and
$\bar \lambda(x)$ in Euclidean space-time. The symmetry transformations
of the lattice field tensor $P_{\mu\nu}(x)$, the operators $\chi(x)$,
$\bar \chi(x)$ and the gauge field in the adjoint representation $V_\mu(x)$
are then easily derived. We also derive useful identities
between the gluino propagator $Q(x;x_0;U_\mu)$, computed on a single
background gauge-field configuration, and the one computed on the
symmetry-transformed background field. The same properties are also derived
for the propagator of the $\tchi$ operator.
Finally, we derive eq.~(\ref{eq:chiff}).

Under parity $\ct{P}$ we have:
\bea
(\vec x, t_0) &\rightarrow& (\vec x^{\ct{P}},t_0^{\ct{P}}) = (-\vec x, t_0)
\nonumber \\
\psi(x) &\rightarrow& \psi^{\ct{P}}(x^{\ct{P}}) = \gamma_0 \psi(x)
\nonumber \\
\bar \psi(x) &\rightarrow& \bar \psi^{\ct{P}}(x^{\ct{P}}) = \bar \psi(x) 
\gamma_0
\nonumber \\
U_0(x) &\rightarrow& U_0^{\ct{P}}(x^{\ct{P}}) = U_0(x)
\nonumber \\
U_k(x) &\rightarrow& U_k^{\ct{P}}(x^{\ct{P}}) = U_{-k}(x) =
U^\dagger_k(x-\hat k)
\nonumber \\
P_{0k}(x) &\rightarrow& P_{0k}^{\ct{P}}(x^{\ct{P}}) = -P_{0k}(x)
\label{eq:par} \\
P_{lk}(x) &\rightarrow& P_{lk}^{\ct{P}}(x^{\ct{P}}) = P_{lk}(x)
\nonumber \\
\chi(x) &\rightarrow& \chi^{\ct{P}}(x^{\ct{P}}) = \gamma_0 \chi(x)
\nonumber \\
\bar \chi(x) &\rightarrow& \bar \chi^{\ct{P}}(x^{\ct{P}}) = \bar \chi(x) 
\gamma_0
\nonumber \\
V_0(x) &\rightarrow& V_0^{\ct{P}}(x^{\ct{P}}) = V_0(x)
\nonumber \\
V_k(x) &\rightarrow& V_k^{\ct{P}}(x^{\ct{P}}) = V_{-k}(x) =
V_k(x-\hat k)
\nonumber
\eea

Under time-reversal $\ct{T}$ we have:
\bea
(\vec x, t_0) &\rightarrow& (\vec x^{\ct{T}},t_0^{\ct{T}}) = (\vec x, -t_0)
\nonumber \\
\psi(x) &\rightarrow& \psi^{\ct{T}}(x^{\ct{T}}) = \gamma_0 \gamma_5 \psi(x)
\nonumber \\
\bar \psi(x) &\rightarrow& \bar \psi^{\ct{T}}(x^{\ct{T}}) = \bar \psi(x) 
\gamma_5 \gamma_0
\nonumber \\
U_0(x) &\rightarrow& U_0^{\ct{T}}(x^{\ct{T}}) = U_{-0}(x) =
U^\dagger_{0}(x-\hat 0)
\nonumber \\
U_k(x) &\rightarrow& U_k^{\ct{T}}(x^{\ct{T}}) = U_{k}(x)
\nonumber \\
P_{0k}(x) &\rightarrow& P_{0k}^{\ct{T}}(x^{\ct{T}}) = -P_{0k}(x)
\label{eq:trev} \\
P_{lk}(x) &\rightarrow& P_{lk}^{\ct{T}}(x^{\ct{T}}) = P_{lk}(x)
\nonumber \\
\chi(x) &\rightarrow& \chi^{\ct{T}}(x^{\ct{T}}) = \gamma_0 \gamma_5 \chi(x)
\nonumber \\
\bar \chi(x) &\rightarrow& \bar \chi^{\ct{T}}(x^{\ct{T}}) = \bar \chi(x) 
\gamma_5 \gamma_0
\nonumber \\
V_0(x) &\rightarrow& V_0^{\ct{T}}(x^{\ct{T}}) = V_{-0}(x) =
V_0(x-\hat 0)
\nonumber \\
V_k(x) &\rightarrow& V_k^{\ct{T}}(x^{\ct{T}}) = V_k(x)
\nonumber
\eea

These two discrete symmetries result in the following useful properties
of the gluino propagator $Q(x;y;U_\mu)$ 
valid on a single gauge-field configuration:
\bea
Q(x;y;U_\mu) &=& \gamma_0 Q(x^{\ct{P}};y^{\ct{P}};U_\mu^{\ct{P}}
(x^{\ct{P}})) \gamma_0 ,
\nonumber \\
Q(x;y;U_\mu) &=& \gamma_5 \gamma_0 Q(x^{\ct{T}};y^{\ct{T}};U_\mu^{\ct{T}}
(x^{\ct{T}}))\gamma_0 \gamma_5 .
\label{eq:propptr}
\eea
The above properites are identical to the ones valid for Dirac fermions.
Also, the usual $\gamma_5$ property, valid for Dirac fermions,
is also true for Majorana fermions:
\bea
Q(x;y;U_\mu) &=& \gamma_5 Q(y;x;U_\mu)^\dagger \gamma_5 .
\label{gamma5}
\eea
We also re-write eq.~({\ref{eq:marco}) in the form
\bea
Q(x;y;U_\mu) &=& C^{-1} Q(y;x;U_\mu)^T C.
\label{C}
\eea
We combine eqs.~(\ref{gamma5}) and (\ref{C}) to obtain:
\bea
Q(x;y;U_\mu) &=& C^{-1} \gamma_5 Q(x;y;U_\mu)^* \gamma_5 C.
\label{marco}
\eea
The transpose superscript $T$ and the Hermitian transpose one
$\dagger$ only act on spin and colour indices, i.e. the transposition 
in space-time is explicit.

We now turn to the two-point correlation function of the $\tchi$ field,
computed on a single background gauge-field configuration, which we denote
by $[C_{\tchi}(t;U_\mu)]_{\alpha\beta}$.
We write
\be
[C_{\tchi}(t;U_\mu)]_{\alpha\beta} = \sum_{\vec x, \vec x_0} 
[Q_{\tchi}(\vec x, t ; \vec x_0, t_0; U_\mu)]_{\alpha\beta}
\ee
in terms of a $\tilde \chi$-propagator $Q_{\tchi}$:
\bea
\label{eq:chiprop}
[Q_{\tchi}(\vec x, t ; \vec x_0, t_0; U_\mu)]_{\alpha\beta} = 
 \frac{1}{4} \sum_{\vec x,\vec x_0}
\left(\sigma_{\mu\nu}\right)_{\alpha\alpha'} P_{\mu\nu}^a(\vec x,t)
Q^{ab}_{\alpha'\beta'}(\vec x, t ; \vec x_0, t_0 ; U_\mu)
P_{\rho\sigma}^b(\vec x_0,t_0)\left(\sigma_{\rho\sigma}\right)_{\beta'\beta} .
\nonumber \\
\eea
We shall drop the spinor subscripts $\alpha$,$\beta$ from now on.
All the properties (\ref{eq:propptr})--(\ref{marco}) are also satisfied 
by $Q_{\tchi}$, as can be found by using eqs.~(\ref{eq:par}), (\ref{eq:trev})
and (\ref{eq:chiprop}).
The $\ct{P}$ and $\ct{T}$ properties of $Q_{\tchi}(x,x_0;U_\mu)$, listed in
eqs.~(\ref{eq:propptr}), imply for the correlation $C_{\tchi} (t;U_\mu)$:
\be
C_{\tchi} (t;U_\mu) = \gamma_0 C_{\tchi} (t;U_\mu^{\ct{P}}) \gamma_0 ,
\;\;\;\; C_{\tchi} (t;U_\mu) = \gamma_5 \gamma_0 C_{\tchi} (T-t;U_\mu^{\ct{T}})
\gamma_0 \gamma_5 , 
\label{eq:trpcchi}
\ee
for a single gauge-field configuration with periodic boundary
conditions in time. In the first of eqs.~(\ref{eq:trpcchi}) we now
average over field configurations and sum over all space. This we do by
$\sum_{\vec x,\vec x_0} \int \ct{D} U_\mu$ on the l.h.s. and
$\sum_{-\vec x,-\vec x_0} \int \ct{D} U^{\ct{P}}_\mu$ on the r.h.s., in order
to obtain
\be
C_{\tchi} (t) = \gamma_0 C_{\tchi} (t) \gamma_0 .
\label{cp}
\ee
Similarly, on the second of eqs.~(\ref{eq:trpcchi}) we average over field
configurations, taking $\int \ct{D} U_\mu$ on the l.h.s. and
$\int \ct{D} U^{\ct{T}}_\mu$ on the r.h.s., in order to obtain
\be
C_{\tchi} (t) = \gamma_5 \gamma_0 C_{\tchi} (T-t) \gamma_0 \gamma_5 .
\label{eq:ctr} 
\ee
The $\gamma_5$ property of eq.~(\ref{gamma5}), combined with translational
invariance of the propagator implies that
\be
C_{\tchi} (t) = \gamma_5 C_{\tchi} (T-t)^\dagger \gamma_5 ,
\label{eq:g5chi}
\ee
whereas from eq.~(\ref{C}) and translational invariance we obtain
\be
C_{\tchi} (t) = -C C_{\tchi} (T-t)^T C.
\label{eq:Cchi}
\ee
Finally, combining eq. (\ref{eq:ctr}) with (\ref{eq:g5chi}), and (\ref{eq:g5chi}) with (\ref{eq:Cchi}), we obtain 
\be
C_{\tchi} (t) = \gamma_0 C_{\tchi}^\dagger (t) \gamma_0, \;\;\;\;C_{\tchi} (t) = - C \gamma_5 C_{\tchi}^* (t) \gamma_5 C. 
\label{cf}  
\ee

Now, let us consider the most general form of the correlation function
$C_{\tchi}$ as a superposition of Dirac matrices:
\be
C_{\tchi}(t) = \sum_i C_i(t) \Gamma_i,
\ee
with $\Gamma_i = I , \gamma_5, \gamma_\mu, \gamma_\mu \gamma_5,
i \sigma_{\mu\nu}$. Using eqs.~(\ref{cp}) and (\ref{cf}), we reduce this expression to
\be
C_{\tchi}(t) =  C_1(t) I + C_2(t) \gamma_0.
\ee
Moreover, from eq.~(\ref{eq:ctr}) we derive that 
\bea
C_1(t) &=& C_1(T-t) , \nonumber \\
C_2(t) &=& - C_2(T-t),
\eea
which, combined with eq.~(\ref{eq:g5chi}), yields that $C_1(t)$ and $C_2(t)$
are real.

By performing analogous manipulations, it is easy to obtain the same
results for the ``spatial" correlation $C_{\tchi}^S (t)$.

\section{Supersymmetric Ward identity}
\label{app:xaxs}


The supersymmetric lattice transformations for the gluino fields are:
\be
  \label{susytr_f}
 \left \{ \begin{array}{lll}
  \delta_\epsilon \lambda(x) &=& - \frac{1}{2} \sigma_{\mu\nu} P_{\mu\nu}
\epsilon  \\ \\
  \delta_\epsilon \bar \lambda(x) &=& + \frac{1}{2} \bar \epsilon \sigma_{\mu\nu} P_{\mu\nu} 
\end{array} \right. 
\ee
where $P_{\mu\nu}(x)$, defined in eq.~(\ref{eq:pmunu}), has dimension 2.
The supersymmetric transformations of the link variables are:
\be
  \label{susytr_g}
  \left \{ \begin{array}{lll}
\delta_\epsilon U_\mu(x) &=& + i a g \bar \epsilon \gamma_\mu \lambda (x) 
U_\mu(x)  \\ \\
\delta_\epsilon U^\dagger_\mu(x) &=& - i a g U^\dagger_\mu(x) \bar 
\epsilon \gamma_\mu \lambda(x) \end{array} \right. 
\ee
In eqs.~(\ref{susytr_f}) and (\ref{susytr_g}), $\epsilon$ is an 
infinitesimal fermionic parameter of dimension $-1/2$. 
These transformations commute with the gauge transformations and give,
in the continuum limit, the continuum SUSY transformation for the 
vector supermultiplet.

The supersymmetric WI is:
\begin{equation}
  \label{susywi}
\nabla_\mu S_\mu(x) = m_0 \sigma_{\rho\sigma} P^a_{\rho\sigma}(x) \lambda^a(x) + X_S(x) .
\end{equation}
The supersymmetric current $S_\mu(x)$ is a dimension 7/2 operator, defined as
\begin{equation}
  \label{susycurr}
  S_\mu(x) = - \sigma_{\rho\sigma} \gamma_\mu P^a_{\rho\sigma}(x)
V^{ab}_\mu(x) \lambda^b(x+\mu) .
\end{equation}
In the limit $a \to 0$ it reduces to the continuum supersymmetric current.

The operator $X_S(x)$ is given by
\begin{eqnarray}
  \label{x_s}
  X_S(x) &=&  \frac{1}{a} \sigma_{\alpha\beta} (\gamma_\mu + r) \left [ P^a_{\alpha\beta}(x) \lambda^a(x) 
                     - \frac{1}{2} P^a_{\alpha\beta}(x) V^{ab}_\mu(x) \lambda^b (x+\mu) \right. \nn \\
&& - \left. \frac{1}{2} \lambda^a(x-\mu)V^{ab}_\mu(x-\mu) P^b_{\alpha\beta}(x) \right ]  \\
&+& \frac{1}{a} \gamma_\nu \gamma_5 \left [ \lambda^a(x) \tilde P^a_{\mu\nu}(x)
      - \tilde P^a_{\mu\nu}(x-\mu) V^{ab}_\mu(x-\mu) \lambda^b(x) \right] \nn \\
&-& g \gamma_\mu \lambda^b(x) \left [ \bar \lambda^a(x) (\gamma_\mu - r) f_{abd}
V^{dc}_\mu(x) \lambda^c ( x + \mu) \right ] , \nn
\end{eqnarray}
where $f_{abc}$ is the structure constant of $SU(2)$. In the naive continuum limit, $X_S$
is proportional to the lattice spacing times a dimension 11/2 composite operator.
$X_S$ does not contain only terms proportional to $r$. This is because
supersymmetry is broken by both the Wilson term and the 
explicit Lorentz group breaking (due to the finiteness of the lattice 
spacing). The operator $X_S(x)$ is {\em irrelevant} at tree level, but
divergent at one loop, since it mixes with operators of dimension 11/2 or
less under renormalization \cite{cv}.

\end{document}